\documentclass[%twocolumn,
reprint,
superscriptaddress,
 amsmath,amssymb,
 aps,pra
]{revtex4-2}

\usepackage{graphicx}%
\usepackage{multirow}%
\usepackage{amsmath,amssymb,amsfonts}%
\usepackage{amsthm}%
\usepackage{mathrsfs}%
\usepackage[title]{appendix}%
\usepackage{xcolor}%
\usepackage{textcomp}%
\usepackage{booktabs}%
\usepackage{listings}%
\usepackage{indentfirst}
\usepackage{float}
\usepackage{hyperref}
\usepackage[all]{hypcap}
\usepackage[normalem]{ulem}
\usepackage{upgreek}
\hypersetup{
    colorlinks=true,
    linkcolor=blue,
    filecolor=blue,      
    urlcolor=blue,
    citecolor=blue
    }

\usepackage{soul} % for striking through text

\newcommand{\beq}{\begin{equation}}
\newcommand{\eeq}{\end{equation}}%revtex4-2.cls
\newcommand{\beqa}{\begin{eqnarray}}
\newcommand{\eeqa}{\end{eqnarray}}

\newcommand{\ee}{\mathrm{e}}
\newcommand{\ii}{\mathrm{i}}

\newcommand{\wlat}{\omega_{\mathrm{lat}}}
\newcommand{\Vlat}{V_{\mathrm{lat}}}
\newcommand{\kL}{k_{\mathrm{L}}}

\newcommand{\lambdaL}{\lambda_{\mathrm{L}}}

\newcommand{\wR}{\omega_{\mathrm{R}}}
\newcommand{\ER}{E_{\mathrm{R}}}
\newcommand{\kB}{k_{\mathrm{B}}}

\usepackage[mathlines]{lineno}% Enable numbering of text and display math
\usepackage{bm}

\begin{document}

\title{Phase-Noise-Induced Heating in Optical Lattices}

%\author{Lithium III}
%\address{Laboratoire Kastler Brossel,    ENS-PSL University, CNRS,  Coll\`ege de France, Sorbonne Universit\'e, 24 rue Lhomond, 75005 Paris, France}

\author{Jean Paul Nohra}
\affiliation{Laboratoire Kastler Brossel, ENS-Universit\'{e} PSL, CNRS, Sorbonne Universit\'{e}, Coll\`{e}ge de France, 24 rue Lhomond, 75005, Paris, France}
\author{Cyprien Daix}
\affiliation{Laboratoire Kastler Brossel, ENS-Universit\'{e} PSL, CNRS, Sorbonne Universit\'{e}, Coll\`{e}ge de France, 24 rue Lhomond, 75005, Paris, France}
\author{Joris Verstraten}
\affiliation{Laboratoire Kastler Brossel, ENS-Universit\'{e} PSL, CNRS, Sorbonne Universit\'{e}, Coll\`{e}ge de France, 24 rue Lhomond, 75005, Paris, France}
\author{Maxime Dixmerias}
\affiliation{Laboratoire Kastler Brossel, ENS-Universit\'{e} PSL, CNRS, Sorbonne Universit\'{e}, Coll\`{e}ge de France, 24 rue Lhomond, 75005, Paris, France}
\author{Tim de Jongh}
\thanks{Present address: JILA, National Institute of Standards and Technology, and Department of Physics, University of Colorado, Boulder, CO 80309, USA}
\affiliation{Laboratoire Kastler Brossel, ENS-Universit\'{e} PSL, CNRS, Sorbonne Universit\'{e}, Coll\`{e}ge de France, 24 rue Lhomond, 75005, Paris, France}
\author{Bruno Peaudecerf}
\affiliation{Laboratoire Collisions Agr\'egats R\'eactivit\'e, UMR 5589, FERMI, UT3, Universit\'e de Toulouse, CNRS, 118 Route de Narbonne, 31062, Toulouse CEDEX 09, France}
\author{Fabrice Gerbier}
\affiliation{Laboratoire Kastler Brossel, ENS-Universit\'{e} PSL, CNRS, Sorbonne Universit\'{e}, Coll\`{e}ge de France, 24 rue Lhomond, 75005, Paris, France}
\author{Tarik Yefsah}
\thanks{Correspondence to be addressed to: \href{mailto:tarik.yefsah@lkb.ens.fr}{tarik.yefsah@lkb.ens.fr}}
\affiliation{Laboratoire Kastler Brossel, ENS-Universit\'{e} PSL, CNRS, Sorbonne Universit\'{e}, Coll\`{e}ge de France, 24 rue Lhomond, 75005, Paris, France}

\date{\today}

\begin{abstract}
We experimentally and theoretically study the origin of heating in optical lattices by disentangling the respective roles of intensity and phase noise depending on the lattice parameters. While intensity noise is widely identified as a major limiting factor, we show that phase noise can become the dominant heating source, especially for light atoms and deep optical lattices. We provide a simple theoretical framework to predict the phase-noise-induced heating from the power spectral density of the laser phase noise, which can be measured experimentally. We show that such predictions can accurately reproduce the measured heating rates of lithium-6 atoms in a triangular lattice. Our approach is readily generalized to other lattice geometries and atomic species.
\end{abstract}

\maketitle

%\linenumbers
\section{Introduction}
\label{sec:intro}

Optical lattices, periodic light structures created by the interference of mutually coherent laser waves, play a key role in the field of ultracold atoms\,\cite{ grynberg2001,greiner2002,bloch2008}. They can be realized with laser light either close to an atomic transition for simultaneous cooling and trapping\,\cite{Hemmerich1993,grynberg2001} or far-off resonance to avoid excessive recoil heating caused by spontaneous emission\,\cite{grimm2000}. In this article, we concentrate on the latter situation, which is the most relevant for quantum gases. Applications of optical lattices to quantum gases are numerous, from optical clocks \cite{campbell2017}, to the preparation and manipulation of atomic wavepackets (see, \textit{e.g.}\,\cite{Weidemuller1995,morsch2006,fedoseev2025}), and the emulation of emblematic models in quantum many-body physics, such as the Hubbard\,\cite{greiner2002,jordens2008,schneider2008},  Harper-Hofstadter\,\cite{aidelsburger2013, miyake2013}, or XY models\,\cite{struck2011}. Optical lattices have also enabled the realization of quantum gas microscopy, allowing one to probe quantum gases at the single-atom level both in lattice systems~\cite{Bakr2009,Sherson2010,cheuk2016,parsons2016,boll2016,Gross2017,Gross2021} and in the continuum~\cite{verstraten2025,dejongh2025,xiang2025,yao2025,dixmerias2025,mortlock2025,daix2026,dixmerias2025a,daix2025}. 

In all these applications, a key requirement is to maintain heating of the trapped atomic gas to a negligible level. Heating occurs for both fundamental and technical reasons. The main fundamental heating mechanism is recoil heating from residual spontaneous emission, which can, in most cases, be mitigated to a manageable level by a large laser detuning. Technical heating mechanisms come mostly from laser noise. Laser-noise-induced heating in optical traps (without interference) was studied in a seminal paper by Savard \textit{et al.}\,\cite{savard1997} (see also \cite{gehm1998,gehm2000,gardiner2000}). They identified two main contributions: (i) intensity fluctuations, which lead to parametric (exponential) heating, and (ii) beam-pointing fluctuations, which give rise to linear heating. In the context of optical lattices, beam-pointing instabilities are generally negligible, and experimental efforts have therefore focused on minimizing intensity noise to realize low-amplitude-noise optical lattices \cite{blatt2015,mazurenko2019}. %More recently, phase-stabilization schemes have also been proposed for time-dependent and tunable optical lattices \cite{mehling2026,aksentsev2025}.

In this Article, we show that the phase noise of the laser used to generate a static optical lattice can be a significant source of heating, and may even become the dominant mechanism for light atomic species or very deep lattices, as commonly employed in quantum gas microscopy. We support this conclusion through measurements of the heating rate of ultracold $^6$Li gases trapped in a two-dimensional optical lattice, and by comparing these measurements with predictions from a Fermi golden rule model using the experimentally measured phase-noise spectrum. Our work provides a predictive framework for selecting lasers for optical lattice experiments and evaluating the needs for active stabilization of the laser intensity and phase.

\section{Heating in fluctuating optical lattices}
\label{Sec:heating_theory}
\subsection{Fluctuating one-dimensional optical lattices}
\label{Sec:1dlattice}

\begin{figure}[t!!!]
\centering
\includegraphics[width=\columnwidth]{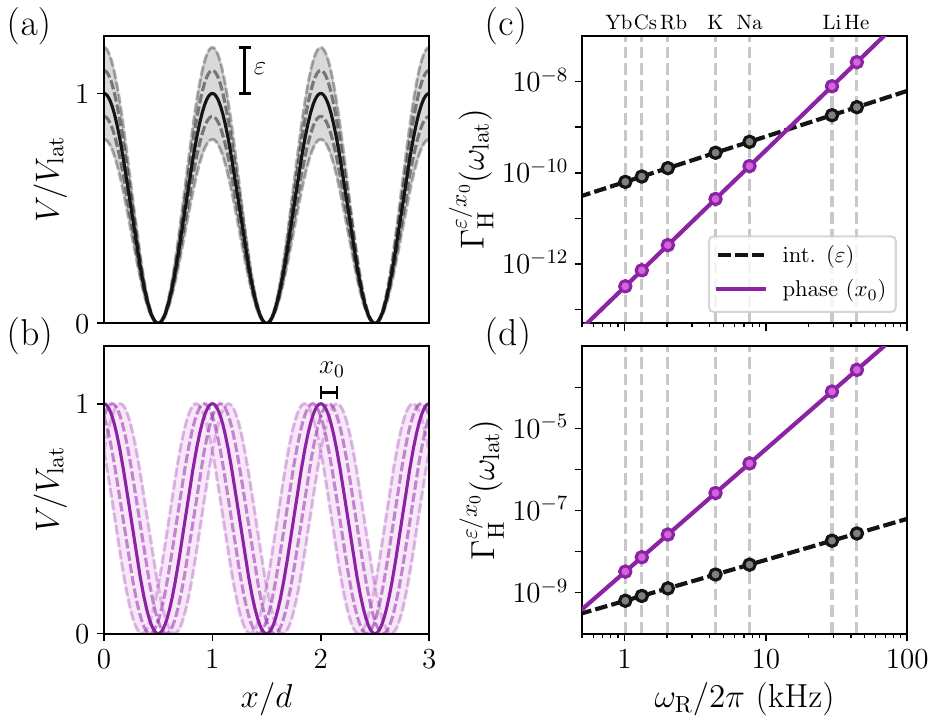}
\caption{ (\textbf{a}) Intensity fluctuations $ \varepsilon=\delta I/\bar{I}$ translate into fluctuations of the amplitude of the lattice potential with spacing $d$, whereas (\textbf{b}) relative phase fluctuation $\phi_{21}  $  translate into a random translation of the potential by $x_{0} =\phi_{21}/\kL$. Both processes result in heating of an atomic gas trapped in said lattice. In the right panel, we compare the normalized heating rates for several atomic species in two regimes of experimental interest characterized by the reduced lattice depth $s = V_{ \mathrm{lat}}/\ER$: (\textbf{c}) $s =10$, corresponding to the Hubbard regime, and (\textbf{d}) $s =1000$, corresponding to the quantum gas microscopy regime.  The black and purple solid lines show the rate of heating caused by intensity or phase fluctuations, respectively. We used the illustrative values $S_{ \varepsilon} \left(2 \wlat  \right) =10^{-15}$\,Hz$^{-1}$,  $S_{\phi_{\mathrm{L}}}\left( \wlat\right)=10^{-9}$\,Hz$^{-1}$ and $\tau=10^{-9}$\,s (see text), with $\lambda_{\rm L}=1064\,$nm.}
\label{figure1}
\end{figure}

We start by examining the simplest situation: a single particle in one dimension trapped in the periodic potential created by a standing wave,
\begin{align}
V(x) & =  V_{ \mathrm{lat}} \big[1+ \varepsilon\big] \sin^2\big[  \kL ( x-x_{0})\big].
 \end{align}
The potential depth $V_{ \mathrm{lat}} = s \ER$ is proportional to the mean intensity $\bar{I}$ of the lattice lasers, where $\ER=\hbar^2 \kL^2/(2M)$ is the recoil energy, $M$ being the atomic mass, and $\kL=2\pi/\lambda_{\rm L}$ is the laser wavenumber, with $\lambda_{\rm L}$ its wavelength. We note $\wR = \ER/ \hbar$ the corresponding recoil frequency. Intensity fluctuations quantified by the relative intensity noise $ \varepsilon(t)=\delta I(t)/\bar{I}$ lead to random modulations of the lattice amplitude. Fluctuations of the relative phase $\phi_{21}$ of the two laser arms creating the standing wave lead to random translation of the lattice potential as a whole by a quantity $x_{0}(t) = \phi_{21}(t)/\kL$ (see Fig.~\ref{figure1}a and b). Both processes induce inter-band transitions and lead to heating of the trapped particle. Note that the fluctuating part of the potential,  because it has the same symmetry as the static part, does not induce quasi-momentum-changing transitions, including intra-band transitions.

The heating is easily quantified in the regime of deep lattices, where the atoms are tightly trapped near the minima $x_{m}$ ($m \in \mathbb{Z}$) of $V(x) $. For $x \approx x_{m}$, the potential is well approximated by the fluctuating harmonic potential
\begin{align}
V(\delta x=x-x_m ) & \approx   \frac{M \wlat^2 }{2}\big[1+ \varepsilon(t)\big]  \big[ \delta x-x_{0}(t)\big]^2,
\end{align}
with $\wlat = 2\wR \sqrt{s}$ the oscillation frequency near the bottom of a lattice well. Heating of a harmonically trapped particle caused by fluctuations of the trap spring constant $\varepsilon(t)$ and center $x_{0}(t)$ has been considered previously in Savard \textit{et al}\,\,\cite{savard1997,gehm1998} in the context of optical dipole traps. Fluctuations of the spring constant originating from intensity fluctuations lead to parametric heating where the mean energy increases exponentially, $\dot{E}/E = \Gamma_{ \mathrm{H}}^{ \varepsilon} $. For the lattice potential considered here, the parametric heating rate follows from the Fermi golden rule,
\begin{align}
\label{eq:gammaH_epsilon}
\Gamma_{ \mathrm{H}}^{ \varepsilon} & = \frac{\pi\wlat^2}{2} S_{ \varepsilon} (2 \wlat),
\end{align} where we denote by
\begin{align}
\label{eq:S_omega}
S_{ \zeta} ( \omega) & = \frac{1}{\pi} \int_{ -\infty}^{ +\infty } \langle \zeta(t) \zeta(t+\tau) \rangle \ee^{  \ii\omega\tau}d\tau
\end{align}
the power spectral density associated with the stationary and centered stochastic process $\zeta(t)$. Fluctuations of the trap center lead instead to linear heating, $\dot{E} = \hbar \wlat \Gamma_{ \mathrm{H}}^{ x_0} $, with a golden rule rate 
\begin{align}
\label{eq:gammaH_x0}
\Gamma_{ \mathrm{H}}^{x_0} & =\frac{\pi M \wlat^3}{2\hbar} S_{ x_0} ( \wlat).
\end{align}

In optical dipole traps, fluctuations in the trap center are caused by beam-pointing instability\,\cite{savard1997}. In optical lattices, another mechanism leading to fluctuations of the trap center arises from phase fluctuations $\phi_{\mathrm{L}}(t)$ of the parent laser beam. Such fluctuations translate into fluctuations of the relative phases of the interfering beams when they propagate along paths with unequal length before crossing. 

To account for this phenomenon quantitatively, we relate the power spectral density $S_{ x_0}$ of the trap center $ x_{0}$ in Eq.~\eqref{eq:gammaH_x0} to the power spectral density $S_{\phi_{\mathrm{L}}}$ of the laser phase fluctuations. In the one-dimensional case, we may write the relative phase $\phi_{21}(t) = \phi_{\mathrm{L}}(t- \tau)-\phi_{\mathrm{L}}(t)$, where $\tau=L/c$ is the time delay due to the optical path difference $L$ between the first and the second laser arm, $c$ being the speed of light. We find that $S_{x_{0}}$ and $S_{\phi_{\mathrm{L}}}$ are related by
\begin{align}
S_{x_{0}}(\omega)  = & \frac{2}{\kL^2}\big[1 - \cos( \omega\tau)\big]  S_{\phi_{\mathrm{L}}}(\omega) .
\label{eq:S_x01D}
\end{align}
For short enough time delays $\omega\tau \ll 1$, the rate in Eq.~\eqref{eq:gammaH_x0} becomes
\begin{align}
\Gamma_{ \mathrm{H}}^{x_0} &  \simeq \frac{\pi  \tau^2  \wlat^5 }{4\wR} S_{\phi_{\mathrm{L}}} ( \wlat).
\label{eq:Gamma_1D}
\end{align}

This is the first main result of this work. We expect the scaling $\Gamma_{ \mathrm{H}}^{x_0}  \sim ( \tau^2  \wlat^5 /\wR) S_{\phi_{\mathrm{L}}} (\wlat)$ to hold in general for any (single wavelength) lattice geometry and in any dimension, with the geometric details only affecting a numerical prefactor of order unity. As shown by the $\tau^2$ scaling, phase-noise-induced heating is expected to increase with the difference in beam path between successive crossings. Note that Eqs.~\eqref{eq:S_x01D}-\eqref{eq:Gamma_1D} are readily generalizable to the case of phase-stabilized optical paths, with $\phi_{21}(t)$ set by the performance of the servo loop.

\subsection{Intensity-induced versus phase-induced heating}

To compare the two heating mechanisms, we consider the normalized rates 
\begin{align}
\frac{\Gamma_{ \mathrm{H}}^{ \varepsilon}}{ \wlat} & \simeq\pi \wR \sqrt{s}  S_{ \varepsilon} \left(4 \wR \sqrt{s} \right),\\
\frac{\Gamma_{ \mathrm{H}}^{x_0}}{\wlat } & \simeq 4 \pi \tau ^2 \wR^3  s^2  S_{\phi_{\mathrm{L}}}  \left( 2\wR \sqrt{s}\right),
\end{align}
which we have rewritten to highlight the dependence on the reduced lattice depth $s$ and on the recoil frequency (or, equivalently, on the atomic species under consideration). Because of the steep dependence of $\Gamma_{ \mathrm{H}}^{x_0}$ on $\wR$ and $s$, phase fluctuations can become the dominant mechanism for light atoms and deep optical lattices. Usually, the ``Hubbard regime'' corresponds to $s \sim 10$, while quantum gas microscopy requires $s \sim 1000$ or more~\footnote{In principle, one could also distinguish a weak lattice regime with $s \ll 1$, the analog of the regime of weakly-bound electrons in solid-state physics. In practice, this regime is not very common in quantum gas experiments, and the single-particle dynamics is in any case almost always irrelevant. We thus omit this regime from our discussion.}. 

We show in Fig.\,\ref{figure1}c,d the normalized rates for both regimes using illustrative values for the power spectral densities and time delay $\tau$. For $s=10$, intensity fluctuations dominate for almost all atoms except the lightest ones (Li and He, in the examples considered). In contrast, for $s=1000$, phase fluctuations become the dominant heating mechanism. While the plot is mostly illustrative, it summarizes the strategy to gauge the level of heating that should be expected with a particular laser device whose noise characteristics $S_{ \varepsilon} (\omega)$ and $S_{\phi_{\mathrm{L}}}(\omega)$ are known. The trends reported in Fig.\,\ref{figure1} represent the main conclusion of this work, which we will confront with experiments with $^6$Li atoms in Section\,\ref{sec:heating_meas}.

%\begin{figure}[ht!!!]
%\centering
%\includegraphics[width=\columnwidth]{figures/figure_PlotGammaH.pdf}
%\caption{ Normalized heating rates for a lattice depth of $s_0 = V_{ \mathrm{lat}}/\ER =20$ (\textbf{a}, Hubbard regime) and 
%$s_0 =200$ (\textbf{b}, quantum gas microscopy regime). The blue dashed and red solid lines show the rate of heating caused by intensity or phase fluctuations, respectively. We used illustrative values $ \wR S_{ \varepsilon} \left(2 \wlat  \right) =10^{-6}$ and   $\wR S_{\phi_{\mathrm{L}}}\left( \wlat\right)=(2\pi)^2 \times 10^{-6}$ to draw the plots. }
%\label{figure2}
%\end{figure}

\subsection{Phase-noise-induced heating in two-and three-dimensional lattices}

The calculations of Section\,\ref{Sec:1dlattice} are readily extended to different lattice geometries in higher dimensions. The case of cubic lattices, created by three mutually perpendicular and independent one-dimensional lattices, is particularly simple, since it reduces to the case of Section\,\ref{Sec:1dlattice} along each direction. We discuss here explicitly the triangular lattice as used in our experiments.
The lattice potential is formed by folding a single laser beam twice such that the relative angle between two consecutive arms of the lattice is approximately 120$^\circ$. In the following calculation, we will assume that the angle is exactly 120$^\circ$ for simplicity. The lattice sites are then located at the positions $\bm{R}_{mn}=\bm{r}_0 + m \bm{e}_1 + n \bm{e}_2$, with $m,n \in \mathbb{Z}^2$ and $\bm{e}_{1/2}$ the primitive vectors of the triangular lattice and with a global shift
\begin{align}
\label{eq:r0_triangular}
\bm{r}_0 &= 
\begin{pmatrix}
\frac{2 \phi_3-\phi_1-\phi_2}{3\kL}\\
\frac{ \phi_2-\phi_1}{\sqrt{3}\kL}
\end{pmatrix} 
\end{align}
controlled by the laser phases $\phi_1,\phi_2,\phi_3$. Similar to the one-dimensional case, the potential for very deep lattices is well approximated by a harmonic one centered at  $  \bm{R}_{mn}$, here with a lattice trapping frequency $\omega_{ \mathrm{lat}} = \sqrt{ \Vlat \kL ^2/M}=\sqrt{2s} \wR$. 

Restricting ourselves to phase noise and using Eq.\,(\ref{eq:r0_triangular}), we find that the power spectral density of $\bm{r}_0$ is anisotropic (see Appendix~\ref{app:calculSxy}), 
\begin{align}
S_{x_0}(\omega_{\rm lat})  \simeq  \, & 3S_{y_{0}}(\omega_{\rm lat})  \simeq \frac{\tau^2  \omega_{\rm lat}^2}{ \kL^2}  S_{\phi_{\mathrm{L}}}(\omega_{\rm lat}) .
\end{align}
The rates in each direction are still given by Eq.\,(\ref{eq:gammaH_x0}), which results in
\begin{align}
\label{eq:gammaH_x0y0}
\Gamma_{ \mathrm{H}}^{x_0} &  \simeq \frac{\pi  \tau^2  \wlat^5 }{4\wR} S_{\phi_{\mathrm{L}}} ( \wlat) = 3 \Gamma_{ \mathrm{H}}^{y_0}
\end{align}
and a total rate

\begin{align}
\label{eq:gammatot}
\Gamma_{ \mathrm{H}}^{\bm{r}_0} \equiv \Gamma_{ \mathrm{H}}^{x_0} + \Gamma_{ \mathrm{H}}^{y_0} \simeq\frac{\pi  \tau^2  \wlat^5 }{3\wR} S_{\phi_{\mathrm{L}}} ( \wlat) ,
\end{align}
where we recover the same scaling as in Eq.~\eqref{eq:Gamma_1D}, only with a different prefactor.

\section{Experimental results}
\label{sec:heating_meas}
 
In the following, we experimentally test the conclusions of Section~\ref{Sec:heating_theory} by comparing two sets of measurements. On the one hand, we have measured the heating experienced by a two-component cold gas of $^6$Li atoms trapped in optical lattices, and on the other hand, we have quantitatively characterized the phase-noise spectra of the lasers used to generate these lattices, which serve as input to our model from Section~\ref{Sec:heating_theory}. We perform both measurements for two different lasers, a ``noisy'' and a ``quiet'' laser, labeled A and B, respectively.

\subsection{Experimental setup}

\begin{figure}[t!!!]
\centering
\includegraphics[width=\columnwidth]{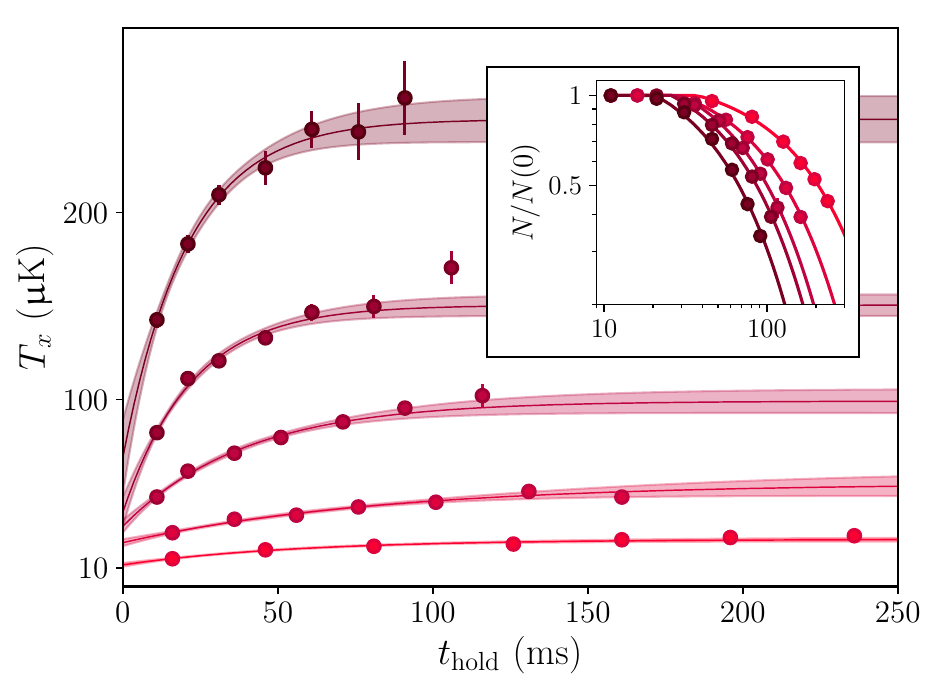}
\caption{Temperature along the $x-$direction versus hold time for a two-component gas of $^6$Li atoms in a triangular optical lattice generated by the noisy laser A. The measurements are shown for representative laser powers, corresponding to trap depths $V_{\rm lat}/\kB = [95(6),159(9), 300(17), 480(20),730(50)]\,\upmu$K from bottom to top. Error bars represent the statistical error from five independent repetitions. The solid lines show the fits to Eq.~\eqref{eq:fit_function} used to extract the heating rates $\dot{Q}_x$, and the shaded areas represent the propagated 1$\sigma$ uncertainty of the fit. The same procedure is applied to the $y-$direction. Inset: corresponding atom number fitted with an exponential decay function in the spilling regime (see text).}
\label{figure2}
\end{figure}

The experiments are performed with a balanced two-component cold gas in the two lowest-lying hyperfine ground states of $^6$Li\,(see \cite{jin2024,verstraten2025} for a detailed description of the experimental apparatus). Briefly, we prepare a nearly degenerate two-component Fermi gas using evaporative cooling in a dipole trap at a bias magnetic field of $B=832\,$ G, where interactions between the two spin components are resonant\,\cite{chin2010}. At the end of the evaporation, we load a two-dimensional optical lattice of variable trap depth $V_{\rm lat}$ with typically  $N\sim6-9\times10^4$ atoms at temperatures $T/T_{\rm F} \sim 2-4$. Here, $T_{\rm F}=\hbar\bar{\omega}_{\rm ini}(6N)^{1/3}$ is the Fermi temperature in the initial trap, with $\bar{\omega}_{\rm ini}$ its geometric mean frequency, $k_{\rm B}$ and $\hbar$ the Boltzmann and reduced Planck constants, respectively. 

The lattice potential has a triangular geometry in the $x-y$ plane, characterized by a lattice spacing $d \simeq 2 \lambdaL/3$, with $\lambdaL = 1064\,$nm the lattice laser wavelength and $\wR/(2\pi) = h/(2M \lambdaL^2) \simeq 29.4\,$kHz the associated recoil frequency. 
The Gaussian envelope of the lattice lasers leads to a nearly harmonic overall potential in three spatial directions, with characteristic frequencies that are at least two orders of magnitude smaller than $
\omega_{\mathrm{lat}}$. This separation of timescales implies that the heating dynamics is two-dimensional, in the sense that we can ignore redistribution of energy along the vertical $z-$direction over the measurement timescale. Furthermore, since tunneling to neighboring sites of the lattice can be safely neglected for the lattice depths used in this work, the presence of the overall trap is irrelevant for the heating dynamics.

At the atoms' location, each lattice arm has a beam waist ($1/e^2$ radius) $w \sim 80\,\upmu$m and a variable power up to $\sim 33\,$W. The beams are derived from laser A in a first set of measurements, and from laser B in a second one. Both lasers are fiber-amplified, single-mode, and single-frequency~\footnote{Manufacturer: Azur Light Systems. Laser A, with reference number ALS-IR-1064-50-I-SF (year 2017), uses an internal seed. Laser B, with reference number ALS-IR-1064-50-A-CP-SF (year 2023), uses the ORION 1064nm laser module from RIO as an external seed.}, but they differ in the laser seed used before amplification, with laser A having a seed with higher phase fluctuations than the seed of laser B. We have measured the phase-noise power spectral density for both lasers using self-heterodyne interferometry (see Appendix\,\ref{app:selfheterodyne}).

 \subsection{Heating  measurements}

 \begin{figure}[t!!!]
\centering
\includegraphics[width=\columnwidth]{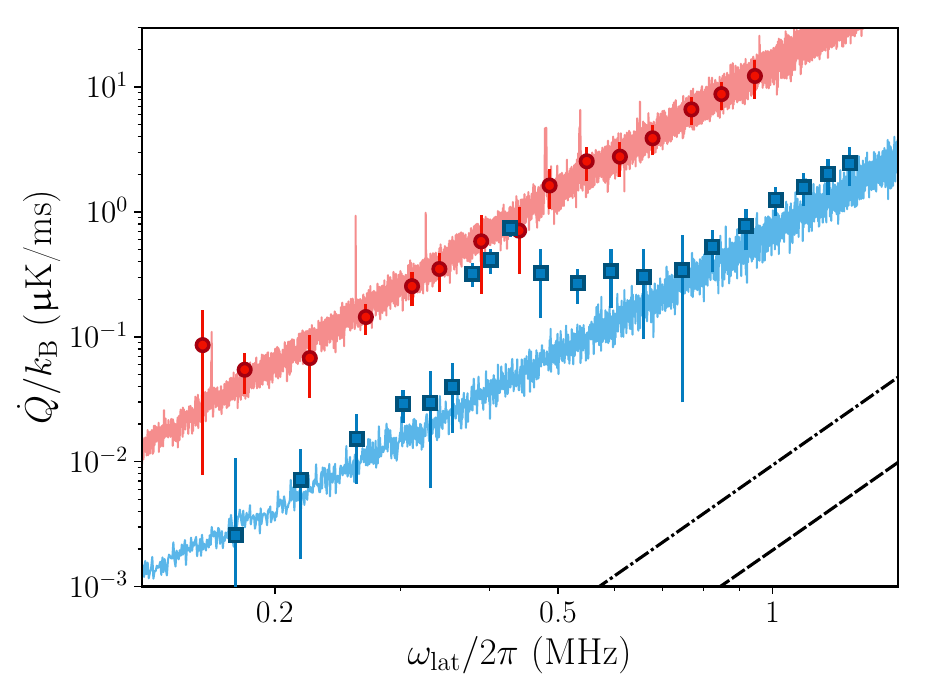}
\caption{Measured heating rate $\dot{Q}=\dot{Q}_x+\dot{Q}_y$ of $^6$Li clouds in optical lattices created using the noisy laser A (red circles) or the quiet laser B (blue squares). The error bars represent the 1$\sigma$ uncertainty propagated from the fitted parameters of Eq.~\eqref{eq:fit_function} (see Figure~\ref{figure2}). The phase-noise-induced heating rates expected from Eq.\,\eqref{eq:gammatot} using the measured phase-noise spectra of lasers A and B are shown by the red and blue traces, respectively. The dashed (resp. dash-dotted) line is the predicted maximal heating rate due to intensity noise, $V_{\rm lat} \Gamma^\varepsilon_{\rm H}$, obtained from Eq.~\eqref{eq:gammaH_epsilon}, with the power spectral density extracted from the specifications of laser A (resp. B).}
\label{figure3}
\end{figure}

After preparation, we hold the atomic gas in the lattice potential for a variable time $t_{\mathrm{hold}}$. We then release the cloud from the traps and record absorption pictures after a variable time of flight up to $800\,\upmu$s (depending on the lattice depth). We extract the total atom number $N$ and the temperature $T$ along each direction from the time-of flight density distributions. Representative curves are shown in Figure\,\ref{figure2}. Two stages are apparent in the evolution. In the first stage, the temperature rises while the atom number remains approximately constant. In the second stage, the temperature levels off while the atom number starts to decrease (inset of Fig.~\ref{figure2}). The behavior beyond the linear regime is qualitatively consistent with atoms eventually acquiring sufficient energy to escape the trap~\cite{gehm1998,gardiner2000}. This spilling regime is of secondary importance for experiments with quantum gases, and we defer its study to Appendix~\ref{app:spilling}. For each direction ($x$ and $y$), we fit the heuristic form
\begin{align}
T(t_{\mathrm{hold}}) =  T_0 + (T_\infty-T_0) \left(1-\ee^{- \gamma t_{\mathrm{hold}}} \right)
\label{eq:fit_function}
\end{align}
to the temperature data to account for the observed saturation of the temperature, with $T_0$, $T_\infty$, and $\gamma$ as free parameters, and extract the heating rates $\gamma (T_\infty-T_0)$. We denote these heating rates, which account for all possible sources of heating in the experiment, with the distinct symbol $\dot{Q}_{x,y}$. Although anisotropic heating could in principle be observed at short times, we observed equal heating rates in both directions within error bars on the time scales at which we probe the system, which we attribute to fast thermalization in the $xy-$plane by two-body collisions.

The measured total heating rates $\dot{Q}=\dot{Q}_x+\dot{Q}_y$ for both lasers are shown in Figure\,\ref{figure3} versus the optical lattice frequency $\wlat$. We compare $\dot{Q}$ to the Fermi golden rule prediction $\dot{E}=\hbar\omega_{\rm lat}\Gamma_{\rm H}^{\bm{r}_0}$, with $\Gamma_{\rm H}^{\bm{r}_0}$ in Eq.\,\eqref{eq:gammatot}, without any free parameters, using the independently measured power spectral density of the lasers (Appendix\,\ref{app:selfheterodyne}) and the known trap frequencies. We also show an estimate of the heating rate due to residual intensity noise (RIN) based on Eq.~\eqref{eq:gammaH_epsilon} and the lasers specifications~\footnote{The manufacturer specifications indicate a RIN ranging approximately from $-140\,$dB/Hz to $-160\,$dB/Hz for laser A, and from $-145\,$dB/Hz to $-153\,$dB/Hz for laser B, in the frequency range $0.1-3\,$MHz relevant to this work.}, which is much lower than the measured heating rates. The expected heating rate due to spontaneous emission (not shown) is comparably low.

For the lattice created with the noisy laser A, the agreement between the heating measurements and the predictions of our model is excellent throughout the entire frequency range. The observed quantitative agreement also validates the assumption that the redistribution of energy towards the $z-$direction is negligible.
Regarding the lattice created with the quieter laser B, phase-noise-induced heating provides a baseline on top of which we observe a broad resonance around $\wlat/(2\pi) \simeq 500\,$kHz. This additional heating is possibly of technical origin, coming from a part of the experimental setup that was not probed in the self-heterodyne measurement. It may, for instance, result from a degraded RIN of laser B compared to the manufacturer's specifications at these frequencies, which we were not able to measure independently. It may also originate from inelastic collisional processes at unitarity (see, e.g.,~\cite{Laurent2017}) in the experiments performed with laser B, where we expect an enhanced three-body recombination rate compared to experiments with laser A, due to peak atomic densities that are about two times higher.

\section{Conclusion}

In conclusion, we have presented a systematic study of phase-noise-induced heating of ultracold atoms in optical lattices. We have shown that phase-noise-induced heating, often considered negligible compared to intensity-noise-induced heating, can in fact become the dominant heating mechanism for light atoms and/or deep optical lattices, such as those employed in quantum gas microscopes. We performed a quantitative comparison between the measured heating rates of $^6$Li gases and the corresponding theoretical predictions of our model, based on independently measured laser phase-noise power spectral densities. The overall good agreement between the two demonstrates that phase-noise-induced heating can be accurately modeled. Our results establish a predictive approach for the selection of laser sources used to generate optical lattices and for evaluating the needs for active stabilization.

\begin{acknowledgments}
This work has been supported by Agence Nationale de la Recherche (Grant No. ANR-21-CE30-0021) and R{\'e}gion Ile-de-France in the framework of DIM QuanTiP.
\end{acknowledgments}
\appendix

\section{Calculation of phase-noise-induced heating for a triangular lattice}
\label{app:calculSxy}

\begin{figure}[ht!!!]
\centering
\includegraphics[width=0.9\columnwidth]{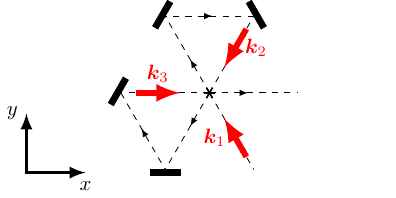}
\caption{ Geometry of the triangular lattice. The incoming beam with wavevector $\bm{k}_1$ is folded twice onto itself to produce the other two beams with wavevectors  $\bm{k}_2$,  $\bm{k}_3$. The triangular lattice potential results from the interference of all three waves at the location of the atoms, marked by an asterisk.}
\label{figure1_SM}
\end{figure}

We consider the lattice geometry shown in Figure\,\ref{figure1_SM}. At the origin (the location of the atoms), the phases of the three beams forming the lattice (indexed 1,2,3)  are related to the phase $\phi_{\mathrm{L}}$ of the incoming laser by  $\phi_{i}(t) = \phi_{\mathrm{L}}[t-(i-1)\tau]$ with $\tau$ the time delay of beam $i$ with respect to beam $i-1$ due to free space propagation. We neglect, for simplicity, any additional noise introduced by vibrations of the folding mirrors and assume a perfectly symmetric geometry.  The correlation functions $C_{x_0} (t) =\big\langle x_{0} (t) x_{0}(0) \big\rangle$ and  $C_{y_0} (t)=\big\langle y_{0} (t) y_{0}(0) \big\rangle $  are then given by
\begin{align}
\nonumber
C_{x_0} (t)  = & \frac{1}{9\kL^2} \Big[ 6 C_{\phi_{\mathrm{L}}} (t) 
- C_{\phi_{\mathrm{L}}} (t+\tau)  -  C_{\phi_{\mathrm{L}}} (t-\tau)  ,\\
&-  2C_{\phi_{\mathrm{L}}} (t +2\tau)-  2C_{\phi_{\mathrm{L}}} (t-2\tau)   \Big],\\
C_{y_0} (t)  = &  \frac{1}{3\kL^2} \Big[ 2 C_{\phi_{\mathrm{L}}} (t) 
- C_{\phi_{\mathrm{L}}} (t+\tau)  
-  C_{\phi_{\mathrm{L}}} (t-\tau)  \Big],
\end{align}
with $C_{\phi_{\mathrm{L}}} (t)  = \big\langle \phi_{\mathrm{L}} (t) \phi_{\mathrm{L}}(0) \big\rangle$ the correlation function of the laser phase. Using the Wiener-Khintchin theorem and the Fourier-transform shift theorem in Eq.~\eqref{eq:S_omega}, we obtain the spectral densities 
\begin{align}
S_{x_0}(\omega)  = &  \frac{2}{9\kL^2} \Big[ 1 - \cos( \omega \tau ) +4\sin^2(\omega \tau )  \Big]  S_{\phi_{\mathrm{L}}}(\omega) ,\\
S_{y_0}(\omega)   = & \frac{2}{3\kL^2} \Big[ 1 - \cos(\omega \tau) \Big] S_{\phi_{\mathrm{L}}}(\omega ) .
\end{align}
In practice, the time delays are very small: for a beam path $L \sim 0.5\,$m, $\tau = L/c \sim 1.6 \cdot 10^{-9}\,$s, and $\omega \tau \leq 10^{-2}$ for typical frequencies $\wlat \leq (2\pi) \times 1\,$MHz. We may then use the ``low-frequency'' limit of the spectral densities quoted in the main text.

 \section{Measurement of the phase noise power spectral density by self-heterodyne interferometry}
\label{app:selfheterodyne}
 
In order to compare the measured heating rates to our model, we determine the power spectral density of the laser phase fluctuations using self-heterodyne interferometry. Briefly, part of the laser beam is split into two arms, one of them including an acousto-optical modulator shifting the beam frequency by $\omega_{\mathrm{c}} \simeq 2\pi \times 80\,$MHz and a 20 m-long optical fiber resulting in an overall propagation delay of $\theta$ between the two arms. After recombining the two beams, a beatnote signal proportional to
$A(t)= A_0  \cos\Phi(t)$, with $\Phi(t) = \phi_{\mathrm{L}}(t-\theta)-\phi_{\mathrm{L}}(t)-\omega_{\mathrm{c}} t$, is measured by a fast InGaAs photodiode (bandwidth 100\,MHz) and recorded on a digital oscilloscope.

\begin{figure}[t!]
\centering 
\includegraphics[width=\columnwidth]{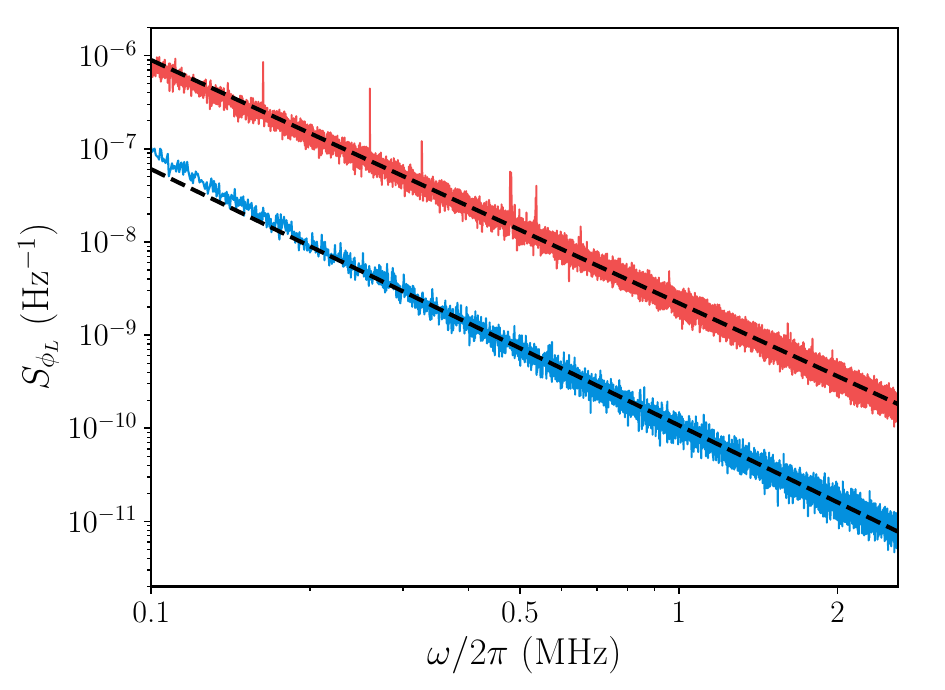}
\caption{Measured power spectral density of the laser phase $S_{\phi_{\mathrm{L}}}(\omega)$ for the noisy laser A (red trace) and the quiet laser B (blue trace). An algebraic fit $\propto\omega^{-\alpha}$ (dashed lines) yields $\alpha=2.60(1)$ and $2.80(2)$ respectively.}
\label{fig_PSD_app}
\end{figure}

We extract the power spectral density $\mathcal{S}_\Phi$ of the interferometer phase $\Phi(t) $ using standard signal processing techniques. We first  calculate the analytic signal  
\begin{align}
\label{eq:analyticsignal}
\widetilde{A}(t)& = A(t) +  \ii \mathrm{H}\Big[A(t)\Big]= A_0  \ee^{\ii \Phi(t)}
\end{align}
associated with $A(t)$, where $\mathrm{H}$ denotes the Hilbert transform. We then obtain the interferometer phase $\Phi(t)$ as the complex argument of $\widetilde{A}(t)$. Note that we assume that the amplitude $A_0$  has negligible fluctuations. In practice, a measurement consists of a digital sample $\{ \Phi(n t_{\mathrm{s}})\}_{n=0,N_{\mathrm{s}}-1}$ recorded with a sampling frequency $f_{\mathrm{s}}=1/t_{\mathrm{s}}\simeq 0.25\,$GHz over a duration $(N_{\mathrm{s}}-1) t_{\mathrm{s}} \simeq  10\,$ms. An estimator $\mathcal{S}_\Phi$ of the true power spectral density $S_\Phi$ (the ``periodogram'') is
\begin{align}
\label{eq:Sestimator}
\mathcal{S}_\Phi(\Omega=\omega_m) & =  \langle \vert \widetilde{\Phi}_T(\omega_m) \vert^2 \rangle,
\end{align}
where $ \widetilde{\Phi}_T$ denotes the discrete Fourier transform of $\Phi_n$ evaluated for $\omega_m = 2\pi m/[(N_{\mathrm{s}}-1) t_{\mathrm{s}}]$,  with $ m \in [0,N_s-1]$, and the brackets $\langle \cdot \rangle = \frac{1}{P} \sum_P \cdots $ denote  averaging over (typically) $P=20$ repeated measurements. 

 Using the Fourier-transform shift theorem, one can show that\,\cite{yuan2022}  
\begin{align}
\label{eq:SphiL}
S_{ \Phi}(\omega) & = 4\omega^2 \sin^2(\omega \theta/2) S_{ \phi_{\mathrm{L}}}(\omega),
\end{align}
with an interferometer transfer function $G(\omega)= 4\omega^2 \sin^2(\omega \theta/2)$.  
The zero of the transfer function $G(\omega)$ at $\omega \theta =2\pi$  in Eq.\,(\ref{eq:SphiL}) allows us to determine precisely the time delay $\theta \simeq 103(2)\,$ns by fitting the envelope in the vicinity of the zero. Using the results of Ref.\,\cite{yuan2022} to correct for the interferometer transfer function and the finite bandwidth, we extract the phase-noise power spectral density shown in Figure\,\ref{fig_PSD_app} from $S_{ \Phi}(\omega)$. The curves shown in Figure\,\ref{fig_PSD_app} are used as input to Eq.\,(\ref{eq:gammatot}) to estimate the expected heating rates in Section\,\ref{sec:heating_meas}. 

%Alternatively, since the time-derivative of the laser phase is by definition the instantaneous frequency, $\dot{\phi}_{\mathrm{L}} = \wL(t)$, we may also report the same results in terms of the \textit{ frequency noise} power spectral density. The relation between $S_{ \Phi}(\Omega)$ and $S_{ \wL}(\Omega)$ is
%\begin{align}
%S_{ \Phi}(\Omega) & = 4\sin^2(\Omega \tau/2) S_{ \wL}(\Omega).
%\end{align}
%The frequency noise power spectral density is shown in Figure\,\ref{fig_PSD_app} for the two different lasers used in the experiments. 

\section{Loss rate in the spilling regime: phenomenological approach} 
\label{app:spilling}

\begin{figure}[t!!!]
\centering
\includegraphics[width=\columnwidth]{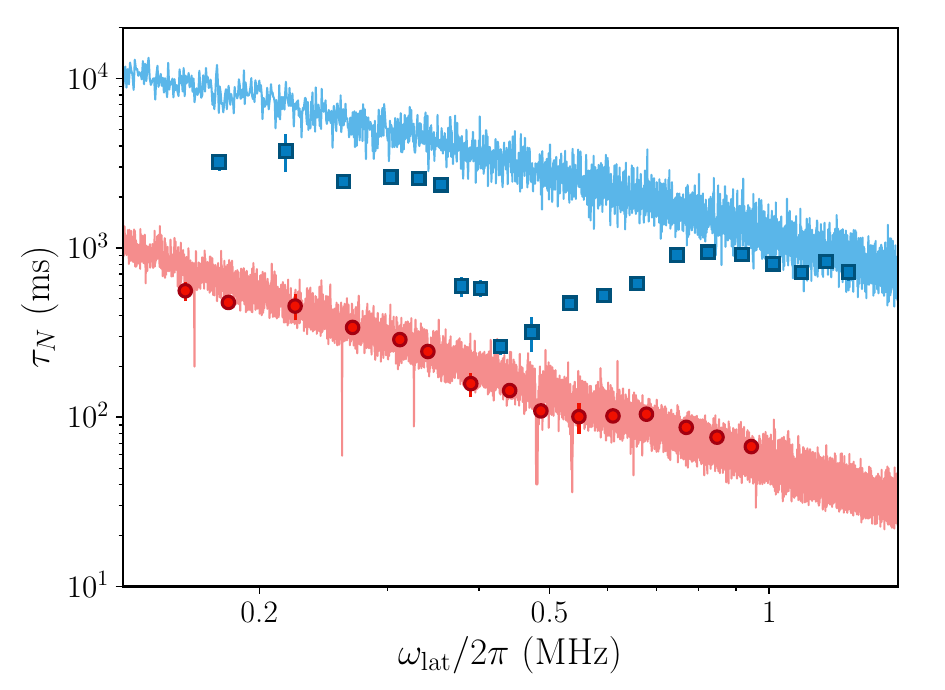}
\caption{Measured lifetime $\tau_N$ of $^6$Li clouds in the spilling regime, for optical lattices created by the noisy laser A (red circles) or the quiet laser B (blue squares). The error bars show the 1$\sigma$ fit uncertainty using the fit function in Eq.~\eqref{eq:lifetime_fit}. The lifetimes expected from Eq.\,\eqref{eq:tau_heuristic} using the measured phase-noise spectra of lasers A and B are shown as red and blue traces, respectively.}
\label{figure_app_Lifetime}
\end{figure}

In the spilling regime, when the temperature becomes comparable to the potential depth, a noticeable fraction of the atoms acquire an energy higher than $V_0$ and escape from the trap (see inset of Figure\,\ref{figure3}). To quantify the loss rate, we fit the atom number by the heuristic function
\begin{align}
N(t)   =&
\begin{cases}
N_0 ,& t < t_{\ast},\\
N_0\ee^{-\frac{t-t_{\ast}}{\tau_{N}} }, & t \geq t_{\ast}.
\end{cases}
\label{eq:lifetime_fit}
\end{align} 
Here the time $t_{_\ast}$ locates the onset of the spilling regime, and $\tau_{N}$ quantifies the lifetime in that regime. The experimental result for $\tau_{N}$ is shown in Figure\,\ref{figure_app_Lifetime}. Also heuristically, we expect that the   lifetime scales as 
 \begin{align}\label{eq:tau_heuristic}
\tau_{N}  \sim \frac{V_{\rm lat}}{\hbar \omega_{\rm lat} \Gamma_{\rm H}^{\bm{r}_0}} = \frac{\sqrt{s/2}}{\Gamma_{\rm H}^{\bm{r}_0}},
\end{align} 
up to a proportionality coefficient. Figure\,\ref{figure_app_Lifetime} compares the measured lifetimes with this expectation (with the proportionality coefficient set equal to one) using the known $V_{\rm lat}$ and the heating rate $\hbar \omega_{\rm lat}\Gamma_{\rm H}^{\bm{r}_0}$ expected from Eq.~\eqref{eq:gammatot} and shown in Fig.~\ref{figure3}. We conclude that this simple approach is sufficient to account for the observed losses (with the previously mentioned caveat regarding the data for laser B near $\wlat/(2\pi)\simeq500$\,kHz). 
 
\newpage

\bibliographystyle{apsrev4-2}
%\bibliography{laser_noise}


%\bibliography{heating_lattice}
%\bibliographystyle{apsrev}

\end{document}